\documentclass{IJMP}
\begin{document}

\title{POSSIBLE $d+\mathrm{i}d$ SCENARIO IN
La$_{2-x}$Sr$_{x}$CuO$_4$ BY POINT-CONTACT MEASUREMENTS}

\author{D. Daghero, R.S. Gonnelli, G.A. Ummarino}

\address{Dipartimento di Fisica, Politecnico di Torino\\
Corso Duca degli Abruzzi 24, 10129 Torino (TO)\\E-mail:
gonnelli@polito.it}

\author{V.A. Stepanov}

\address{Lebedev Physical Institute,
Russian Academy of Sciences \\
Leninski Pr. 53, 119991 Moscow, Russia}

\maketitle

\abstracts{We analyze the results of point-contact measurements in
La$\ped{2-x}$Sr$\ped{x}$CuO$\ped{4}$ (LSCO) previously reported
as a clear evidence of the separation between gap and pseudogap
in this copper oxide. Here we show that, in addition to this, the
conductance curves of our point-contact junctions -- showing
clear Andreev reflection features -- can be interpreted as
supporting a nodeless $d\ped{x^2-y^2}+\mathrm{i}d\ped{xy}$-wave
symmetry of the gap in LSCO. The results of our analysis, in
particular the doping dependence of the subdominant $d_{xy}$ gap
component, are discussed and compared to the predictions of
different theoretical models.}
\section{Introduction}
In spite of the large number of experimental evidences and
theoretical arguments supporting a pure $d_{x^2-y^2}$ symmetry of
the order parameter in cuprates\cite{Tsuei,Brandow}, the possible
existence of a subdominant component with different symmetry has
also been deeply investigated. One of the reasons is that most of
the experimental probes cannot really exclude the presence of a
small additional component. Another reason is that some tunneling
experiments along the $ab$ plane of YBa$_2$Cu$_3$O$_{7-\delta}$
(YBCO) have shown a splitting of the zero-bias conductance peak
(ZBCP) both in the presence\cite{Covington,Dagan,Aprili,Krupke}
and in the absence\cite{Covington,Dagan,Geerk} of a magnetic
field. A possible explanation of this phenomenon stems from the
idea that the ZBCP is due to zero-energy Andreev bound states at
the surface\cite{Hu,Tanaka} that experience a Doppler-like shift
to finite energy in the presence of supercurrents. In the absence
of a magnetic field, such a shift might be due to spontaneous
supercurrents due to the breaking of the time-reversal symmetry.
According to F$\mathrm{\ddot{o}}$gelstrom \emph{et
al.}\cite{Fogelstrom} a subdominant pairing interaction with
smaller critical temperature can in fact appear at the surface of
a $d$-wave superconductor, with a phase shift of $\pi/2$ with
respect to the dominant one. This gives rise to spontaneous
supercurrents and to a local breaking of the time-reversal
symmetry.

An alternative picture has been emerging in the last years, in
which an intrinsic instability of the $d$-wave superconductor
toward a time-reversal breaking state is supposed, with no
relation to surface effects. This picture is somehow based on the
indications of a quantum critical point (QCP) in the proximity of
optimal doping, obtained in Bi$_2$Sr$_2$CaCu$_2$O$_{8+\delta}$
(BSCCO) by ARPES\cite{Valla}. The hypothesis has been made that
such a quantum critical point could mark the transition from a
pure $d$-wave superconducting state to a time-reversal symmetry
breaking state, such as $d_{x^2-y^2}\pm\mathrm{i}s$ or
$d_{x^2-y^2}\pm\mathrm{i}d_{xy}$\cite{Vojta,Khveshchenko,Sachdev,Sangiovanni}.
Recent tunneling data in YBCO at different doping
levels\cite{Dagan} have given some support to this second point
of view, showing that the spontaneous splitting of the ZBCP only
occurs above optimum doping.

In the present paper, we present a possible indication of a
$d+\mathrm{i}d$ scenario in La$_{2-x}$Sr$_{x}$CuO$_4$ (LSCO)
obtained by (re)analyzing the results of point-contact
measurements in polycrystalline LSCO samples with various doping
contents. These data were already reported in a previous
paper\cite{EPJ} in a rather different groundwork, i.e. they were
shown to evidence the separation between superconducting gap and
pseudogap in underdoped LSCO.

\section{Experimental details}
We used La$_{2-x}$Sr$_x$CuO$_4$ polycrystalline samples with
various doping contents from strongly underdoped to slightly
overdoped: $x=$ 0.08, 0.10, 0.12, 0.13, 0.15 and 0.20. Details
about the sample preparation and characterization are given
elsewhere\cite{EPJ,IJMP}. The critical temperatures, determined
by means of magnetic (a.c. susceptibility) and transport
(resistivity) measurements, resulted in good agreement with the
standard $T_{\mathrm{c}}$ vs $x$ curve for LSCO\cite{Takagi}.

Point contacts were obtained by gently pressing sharp Au tips
(whose ending-part diameter was always less than $\sim 2 \; \mu$m)
against the surface of the samples. We often obtained SN
junctions with clear Andreev reflection characteristics. In some
cases, the stability of the point contacts allowed us to follow
the evolution of the conductance curves on heating the junction
from 4.2~K up to the temperature $T_{\mathrm{c}}^{\mathrm{A}}$ at
which the dynamic conductance d$I$/d$V$ was flat.

A discussion of the regime of current flow through our point
contacts was already reported elsewhere\cite{EPJ}. Here let us
just remind that we systematically rejected all the data sets
showing an anomalous temperature and voltage dependence of the
normal-state conductance (for example, for $V>20$ mV) that usually
indicate the presence of heating effects in the
junction\cite{heating}. As a result, all the curves that have
been used for the following analysis can reasonably be thought of
as obtained in a regime of ballistic current flow through the
junction, thus allowing us to perform spectroscopic measurements
with a good energy resolution ($< 1$~meV).

\begin{figure}[t]
\vspace{-3mm}
\begin{center}
\includegraphics[keepaspectratio,width=0.6\textwidth]{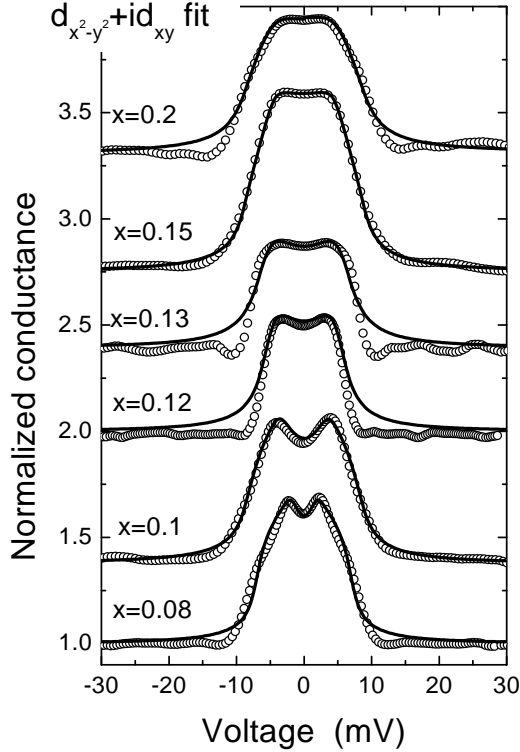}
\end{center}
\vspace{-6mm}\caption{Fit of the conductance curves obtained at
low temperature (4.2 $\div$ 5.6 K) in samples with different
doping levels. The value of the Sr content, $x$, is indicated
near each curve. Symbols: experimental data; solid lines:
best-fitting curves.}\label{fig:d+id_fit}
\end{figure}

\section{Results and discussion}
Fig.~\ref{fig:d+id_fit} reports the experimental conductance
curves at low temperature (4.2 $\div$ 5.6 K) for all the
aforementioned doping contents, normalized to the normal-state
conductance -- so that they tend to unity at high positive
(negative) voltage. The curves have been shifted vertically for
clarity. Solid lines represent the best-fitting theoretical
curves calculated by using the  BTK model\cite{BTK} generalized by
Tanaka and Kashiwaya\cite{Tanaka,TK} with a
$d_{x^2-y^2}+\mathrm{i}d_{xy}$ symmetry of the order parameter.
The details of the fitting procedure are reported
elsewhere\cite{IJMP}. The fitting parameters are
$\Delta_{x^2-y^2}$ and $\Delta_{xy}$, $Z$ (related to the height
of the potential barrier) and the broadening parameter $\Gamma$
that was always kept as small as possible.

Even at a first glance, the fit appears rather good. Notice that
the ``dip'' present in some curves, which is a fairly typical
feature, cannot be fitted at all by the model, irrespective of
the gap symmetry used. It must be said here that, as previously
reported\cite{EPJ}, various other symmetries were tried: $s$,
$d$, $s+\mathrm{i}d$, \emph{extended s} and \emph{anisotropic s},
and none of these could give good results, especially when the
temperature evolution of the curves was considered. Only the
$s+d$ symmetry was found to fit almost equally well the
experimental data \footnote{Actually, for some doping values (e.g.
$x=0.08$, $x=0.15$) the $d+\mathrm{i}d$ fit is considerably
better.}, but its compatibility with the symmetry of the LSCO
lattice at these doping levels is questionable\cite{Tsuei,Klemm}.

\begin{figure}[ht]
\vspace{-3mm}
\begin{center}
\includegraphics[keepaspectratio,width=0.75\textwidth]{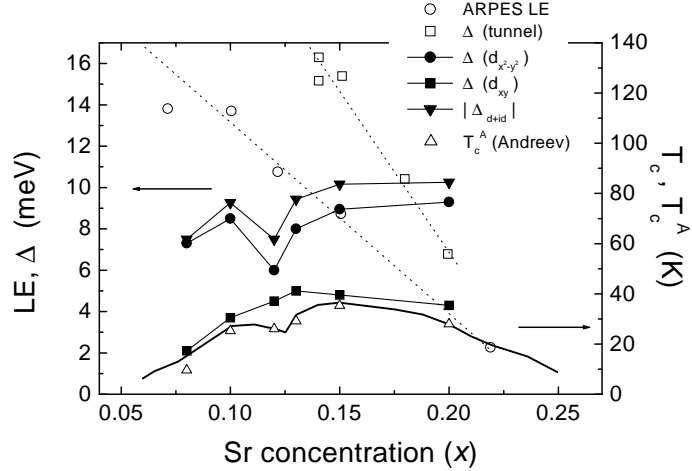}
\end{center}
\caption{Doping dependence of the gap components and of the gap
amplitude, obtained from the fit of the curves in
Fig.~\ref{fig:d+id_fit}. A comparison is made with the results of
tunnel and ARPES  measurements (from refs.[23] and [24],
respectively), and with the standard T$_{c}$ vs $x$ curve (from
ref. [18]) .}\label{fig:gap_vs_x}
\end{figure}

The fit of the low-temperature conductance curves shown in
Fig.~\ref{fig:d+id_fit} gives the doping dependence of the
low-temperature gap components, $\Delta_{x^2-y^2}$ and
$\Delta_{xy}$, reported in Fig.~\ref{fig:gap_vs_x} (solid circles
and squares, respectively). The error affecting each gap value is
rather small (about the size of the points)\footnote{Although
there are 4 fitting parameters, changing each of the gaps has a
very different effect on the curve, and thus the allowed range of
gap values is smaller than expected.}. The amplitude of the gap,
$|\Delta| = \sqrt{\Delta_{x^2-y^2}^{2}+\Delta_{xy}^{2}}$ is also
shown (solid triangles). It is clearly seen that the $d_{xy}$
component is present for all doping levels and is always smaller
than the $d_{x^2-y^2}$ one -- though representing a substantial
part of the total amplitude. Neither $\Delta_{x^2-y^2}$ nor
$\Delta_{xy}$ increase monotonically with decreasing doping, as
instead do both the tunneling gap (open squares) and the ARPES
leading--edge shift (open circles). Rather, a decreasing tendency
is evident in the underdoped region. A comparison is also made
with the standard curve of T$_{c}$ versus doping (thick solid
line)\cite{Takagi}, which is strikingly similar to the
$\Delta_{xy}(x)$ curve and, with less accuracy, to the
$|\Delta|(x)$ one. Notice that a strong suppression of both
$\Delta_{x^2-y^2}$ and $|\Delta|$ occurs at $x=1/8$, where also
T$_{c}$ is reduced, further indicating a close relationship
between the Andreev gap and the critical temperature. Thus, the
conclusion holds true that we drew in a previous paper\cite{EPJ}:
Andreev reflection does measure the superconducting gap, as
opposed to ARPES and tunnel spectrocopies that instead measure the
pseudogap.

\begin{figure}[ht]
\vspace{-3mm}
\begin{center}
\includegraphics[width=0.7\textwidth]{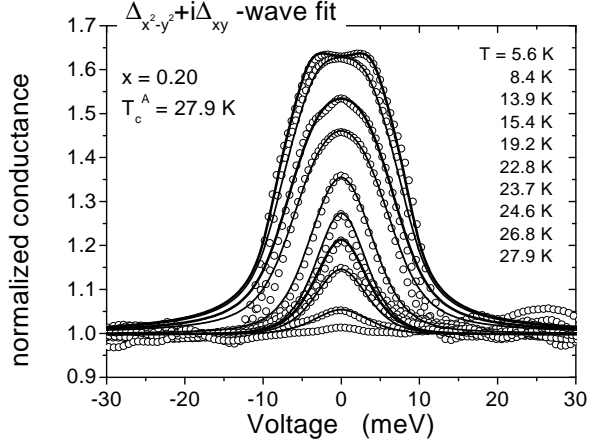}
\end{center}
\vspace{-3mm} \caption{Experimental normalized conductance curve
(symbols) obtained in slightly overdoped LSCO at various
temperatures, from 4.2 K up to the temperature T$_{c}^{A}$ at
which the Andreev-reflection feature disappear together with their
best-fitting curves (lines) calculated by using the generalized
BTK model with $d+\mathrm{i}d$ pairing.
}\label{fig:temperature_fit}
\end{figure}

Further support to this assertion comes from the temperature
dependence of the conductance curves of our junctions. In all
cases, in fact, the Andreev-reflection features disappear at a
temperature T$_{c}^{A}$ close to or smaller than the bulk critical
temperature measured by resistivity, with no evidence of
persistence of the gap above T$_{c}$. The values of T$_{c}^{A}$
are reported for each doping level in Fig.~\ref{fig:gap_vs_x}
(open triangles). Fig.~\ref{fig:temperature_fit} shows, as an
example, the temperature evolution of the curve for $x=0.20$
already shown in Fig.~\ref{fig:d+id_fit}, together with the
$d+\mathrm{i}d$ best-fitting curves obtained by keeping $Z$
constant ($Z=0.135$, which is the value at 4.2~K).

\begin{figure}[ht]
\vspace{-0.8cm}
\begin{center}
\includegraphics[width=0.7\textwidth]{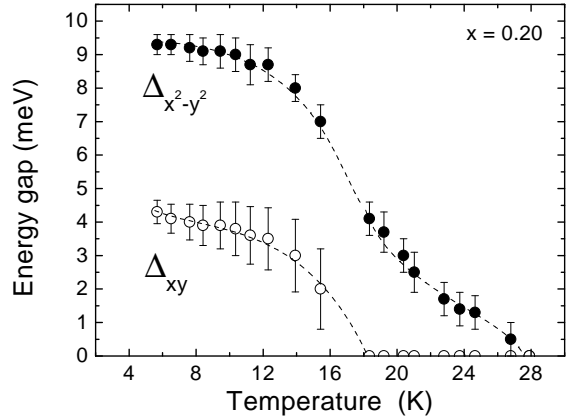}
\end{center}
\vspace{-3mm} \caption{Dependence of the dominant ($d_{x^2-y^2}$)
and subdominant ($d_{xy}$) gap components on the temperature,
obtained from the fit of the conductance curves shown in
fig.\ref{fig:temperature_fit}. Error bars indicate the range of
values that give an acceptable fit when the remaining parameters
are suitably adjusted (note that $Z$ was fixed to its low-$T$
value). Dashed lines are guides to the eye. It is well clear that
the $d_{xy}$ component closes at a lower temperature. The same
happens at all doping contents.}\label{fig:gap_vs_T}\vspace{-3mm}
\end{figure}

Fitting the normalized conductance curves at all temperatures
allows us to obtain the temperature dependence of the two gap
components, which is shown for the case $x=0.20$ in
Fig.~\ref{fig:gap_vs_T}. It is clearly seen that the $d_{xy}$
component is always smaller than the $d_{x^2-y^2}$ one, and that
the thermal evolution of both components follows a very similar
trend, rather different from a BCS curve. Notice that the critical
temperature T$_{c2}$ of the subdominant $d_{xy}$ component is
smaller than T$_{c}$. A very similar thermal evolution of the gap
components is observed also for the remaining doping levels.
Further details will be given in a more extended paper.

\section{Conclusions}
As far as the gap symmetry is concerned, our findings agree with
some tunneling measurements in optimally-doped LSCO, that
evidenced the absence of nodes in the gap\cite{Ekino} and also
with previous Andreev reflection experiments\cite{Achsaf} that
were interpreted as supporting a mixed symmetry. Of course, the
question whether the additional $d_{xy}$ component arises from
surface effects or from a quantum phase transition cannot be
addressed by our measurements. However, it must be said that the
presence of the subdominant $d_{xy}$ pairing in the \emph{whole}
doping range analyzed, as well as its dependence on the doping
(see Fig.~\ref{fig:gap_vs_x}) disagree with the findings in YBCO
films\cite{Dagan}. In that case, the spontaneous splitting of the
zero bias in the tunneling conductance (proportional to the
amplitude of the $d_{xy}$ component) was observed only above
optimal doping, and turned out to increase monotonically with
increasing doping. This behaviour was indeed used to argue for a
quantum critical point near optimal doping, and was reproduced by
some theoretical models predicting the stability of the
$d+\mathrm{i}d$ phase in the overdoped
regime\cite{Sachdev,Sangiovanni}. What our results say, instead,
is that either the time-reversal symmetry breaking is a surface
effect with no relationship to quantum
criticality\cite{Fogelstrom} (and perhaps related to doping only
through the amplitude of the dominant gap component), or the
quantum critical point is placed somewhere in the extreme
underdoped or extreme overdoped region of the phase diagram.
Further measurements in these two extreme regimes will possibly
help in discriminating between these two possibilities.

\vspace{5mm} \noindent{V.A.S. acknowledges the support from RFBR
(project N. 02-02-17133).}

\end{document}